\documentclass[10pt,preprint]{aastex}

\input psfig.sty

\slugcomment{ApJL, in press}

\shorttitle{Formation of gas giant planets}
\shortauthors{Rice \& Armitage}

\begin{document}

\title{On the formation time scale and core masses of gas giant planets}

\author{W.K.M. Rice\altaffilmark{1} and Philip J. Armitage\altaffilmark{2,3}}
\altaffiltext{1}{School of Physics and Astronomy, University of St Andrews, North Haugh KY16 9SS, UK; wkmr@st-andrews.ac.uk}
\altaffiltext{2}{JILA, Campus Box 440, University of Colorado, Boulder CO 80309; pja@jilau1.colorado.edu}
\altaffiltext{3}{Department of Astrophysical and Planetary Sciences, University of Colorado, Boulder CO 80309}

\begin{abstract}
Numerical simulations show that the migration of growing planetary 
cores may be dominated by turbulent fluctuations in the protoplanetary disk, 
rather than by any mean property of the flow. We quantify the impact 
of this stochastic core migration on the formation time scale and core mass 
of giant planets at the onset of runaway gas accretion. For standard Solar Nebula conditions, 
the formation of Jupiter can be accelerated by almost an 
order of magnitude if the growing core executes a random walk with an 
amplitude of a few tenths of an au. A modestly reduced surface density 
of planetesimals allows Jupiter to form within 10~Myr, with an initial core mass below 
10~$M_\oplus$, in better agreement with observational constraints. For 
extrasolar planetary systems, the results suggest that core accretion 
could form massive planets in disks with lower metallicities, 
and shorter lifetimes, than the Solar Nebula. 
\end{abstract}

\keywords{accretion, accretion disks --- solar system: formation ---
	planetary systems: formation --- planets and satellites: individual: Jupiter}

\section{Introduction}
The core accretion model \citep{mizuno80,pollack96} provides the most popular 
explanation for the origin of the Solar System's gas giants, and is consistent 
with the higher frequency of extrasolar planets found to be orbiting metal-rich 
stars \citep{laughlin00,murray02,santos03}. In the simplest version of this 
model, a core grows at a fixed orbital radius from the binary accretion of 
solid planetesimals \citep{safranov69}. Initially, this core is surrounded 
by a near-hydrostatic gaseous envelope, with most of the luminosity being 
provided by ongoing planetesimal accretion. Growth continues until a 
critical mass is exceeded. Above the critical mass there is no stable 
core-envelope solution, and more rapid accretion of the bulk of the 
planetary envelope ensues. 

Observational constraints from the Solar System pose two possible 
problems for core accretion models. First, although Jupiter can  
form within the lifetimes of protoplanetary disks 
\citep{haisch01}, it is hard to form Uranus and Neptune in their present 
locations rapidly enough. This has prompted suggestions that the outer 
giant planets may have migrated outward from birthplaces closer to the 
Sun \citep{thommes99}. Second, upper limits to the  
core mass of Jupiter, derived from Galileo data, are {\em smaller} 
than most theoretical estimates. \cite{guillot99} obtains a firm 
constraint of $M_{\rm core} \le 14 \ M_\oplus$, which is reduced 
to $10 \ M_\oplus$ using a more model-dependent approach. This is 
only marginally consistent with the $10 - 30 \ M_\oplus$ 
core predicted by \cite{pollack96}, and has led to renewed interest 
in models for massive planet formation via 
disk instability \citep{boss97}.

The possibility that orbital migration \citep{goldreich80} of 
planetary cores might reduce the accretion time scale and 
ameliorate these problems was recognized 
by \cite{hourigan84}. In a laminar disk flow, however, gravitational 
torques from the disk induce a rapid, uniformly inward drift \citep{ward97}. 
The benefit of a higher accretion rate must therefore be balanced 
against the reduced residence time of the planet in the disk. In this 
Letter, we point out that this trade-off may not be necessary in 
a turbulent disk. Numerical simulations \citep{laughlin03,nelson03a} 
of the migration of low mass planets within turbulent magnetized 
disks (which extend work by Nelson \& Papaloizou 2003b; Winters, 
Balbus \& Hawley 2003) show that for low enough masses, the 
{\em sense} as well as the rate of 
migration is determined by turbulent 
fluctuations in the disk. As a result, the planet random 
walks in orbital radius as it grows. This behavior was suggested 
as a likely consequence of migration in a magnetized disk by 
\cite{terquem03}. Here, we quantify how 
random walk migration affects massive planet formation.

\section{Method and Assumptions}
We begin our calculations by assuming that a $0.1 M_\oplus$ solid core, 
with density $\rho_{\rm core} = 3.2$ g cm$^{-3}$, has formed at $5$ au in a disk 
with a local surface density of planetesimals of $\sigma_p = 10$ g cm$^{-2}$, and a surface density profile of
$\sigma_p \propto r^{-3/2}$. According to \cite{safranov69} the solid core grows
at the rate $\dot{M}_{\rm core} = \pi R_c^2 \sigma_p \Omega F_g$, where 
$R_c$ is the effective, or capture, radius, 
$\Omega$ is the angular frequency, and $F_g$ is the gravitational enhancement factor \citep{greenzweig92}. 
We will denote the total planet mass (core plus envelope) by $M_p$.

The planetesimal accretion produces a core luminosity of $L_{\rm core} = G M_{\rm core}
\dot{M}_{\rm core}/R_{\rm core}$. With this we can construct quasi-equilibrium models of the envelope
using standard stellar structure equations. The equations of hydrostatic equilibrium, 
mass conservation, and radiative transfer are \citep{papaloizou99,hansen94}
\begin{eqnarray}
\frac{d P}{d R} &=& -g \rho  \\
\frac{d M}{d R} &=& 4 \pi R^2 \rho \\
\frac{d T}{d R} &=& \frac{- \nabla G M \rho T}{R^2 P}.
\label{stellarstruc}
\end{eqnarray}
Here $P$ is the pressure, $\rho$ the density, $T$ the temperature, 
$R$ the distance from the planet center, $g$ the acceleration due to gravity, $M$ the
mass enclosed within $R$, and $\nabla$ the smaller of
the radiative or adiabatic temperature gradients. The inner boundary is located at the core radius and, 
as in \cite{pollack96}, the outer boundary
is located at the smaller of the Hill radius, $R_{\rm Hill} = a (M_p/3M_{\rm star})^{1/3}$,
or the accretion radius, $R_a = G M_p/c_s^2$, where $a$ is the planet's orbital radius, and 
$c_s$ is the local gas sound speed in the disk.
The envelope is taken to be a hydrogen and helium mixture, with mass fractions of 0.7 and 0.28 
respectively, and with an equation of state given by \cite{chabrier92}. Opacities 
from \cite{bell94} are used in calculating the radiative temperature gradient. The 
adiabatic temperature gradient is 
found from the equation of state tables \citep{chabrier92}.

To solve equations (1-3) we need to calculate the planetesimal accretion rate, which 
depends on $\sigma_p$, $F_g$, and $R_c$. 
The planetesimal surface density -- which is more precisely the time-dependent mean surface 
density within the planet's feeding zone -- and the gravitational enhancement 
factor both depend only on $M_p$, and can be calculated using 
expressions given by \cite{pollack96} and \cite{greenzweig92}. For the capture radius we 
adopt a simpler approach than \cite{pollack96}, and take $R_c$ to be
the radius at which a planetesimal with a radius of $100$~km, falling radially through the planet's
envelope, has intercepted a mass of $1/10$ of its own mass. Since this depends upon the 
planet's structure, we iterate at each timestep to find a self-consistent solution. If more than one solution 
is possible (with different planet masses), we choose between them by requiring 
that the total planet mass cannot decrease with increasing time. We then advance 
to the next timestep by incrementing the core mass, and by updating $\sigma_p(r)$ 
to account for the mass of planetesimals accreted 
on to the core in the current timestep. We neglect both the shepherding effect of a 
migrating core upon the planetesimals \citep{tanaka99} (which would reduce the 
accretion rate), and the possibility of diffusion of the planetesimals due to the 
same fluctuating potential that causes core migration (which might increase the 
accretion rate).

\section{Results}
\begin{figure}
\plotone{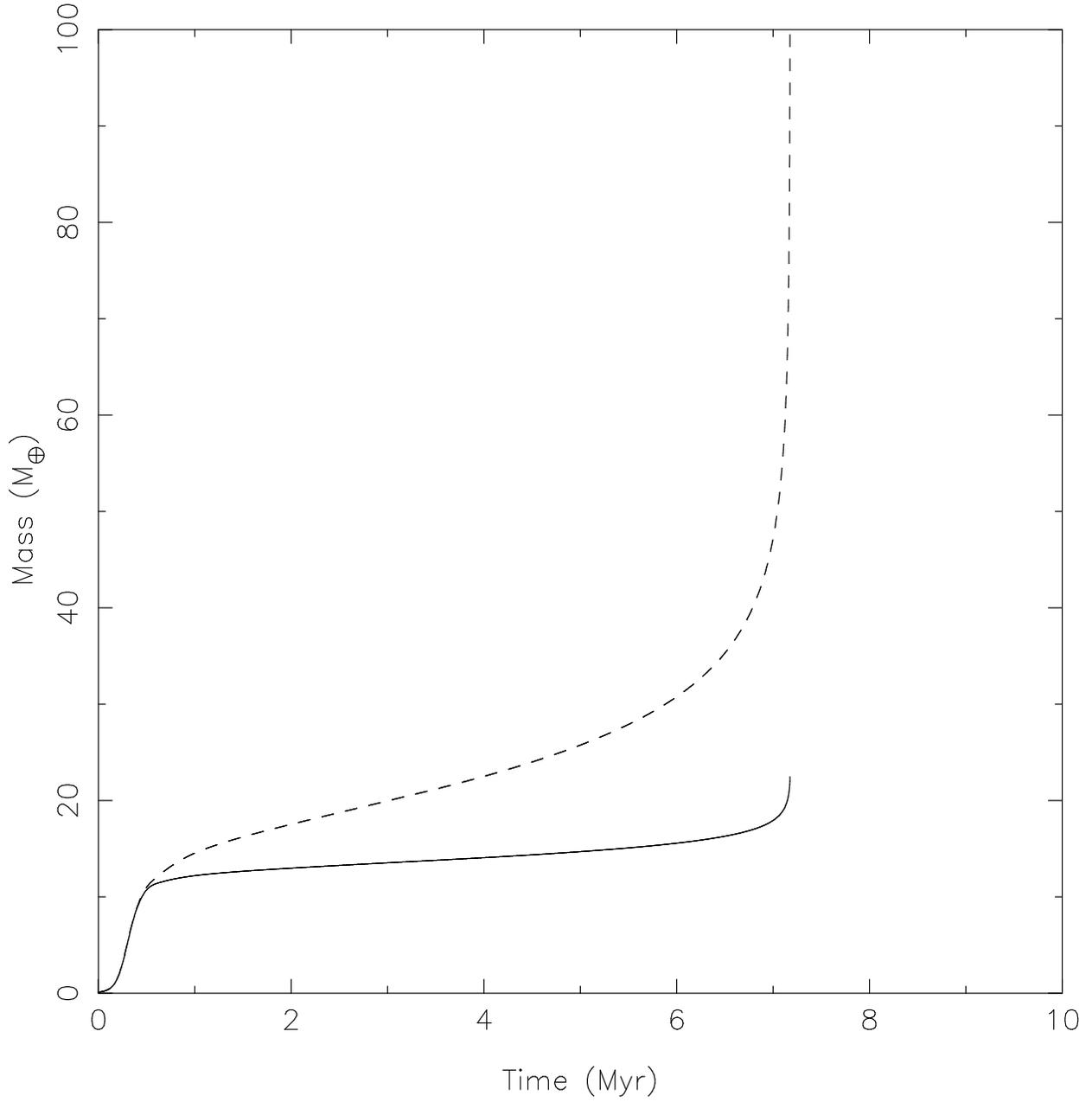}
\caption{Core mass (solid line) and total planet mass
(dashed line) for a core fixed in radius at $5$ au in a 
disk with $\sigma_p = 10 \ {\rm g \ cm}^{-1}$.}
\label{Polmass}
\end{figure}

Our model for planetary growth is a slightly simplified version of that used by 
\cite{pollack96}. If we assume, as they did, that the core remains at a fixed radius, 
almost identical growth rates are obtained. Figure \ref{Polmass} shows the core mass 
and total planet mass against time for a core fixed in radius at $5$ au. There is an 
initially rapid growth phase (their Phase I) in which the core grows to a mass 
of $\sim 10 M_\oplus$ in less than $1$ Myr, followed
by a second phase (Phase II) in which the core grows slowly to a mass of 
$\sim 17 M_\oplus$. Once the core reaches this mass, 
the gaseous envelope grows extremely rapidly (Phase III), 
producing a final planet mass, in this case, of $\sim 100 M_\oplus$, and a total growth
time of just over $\sim 7$ Myr. 

\begin{figure}
\plottwo{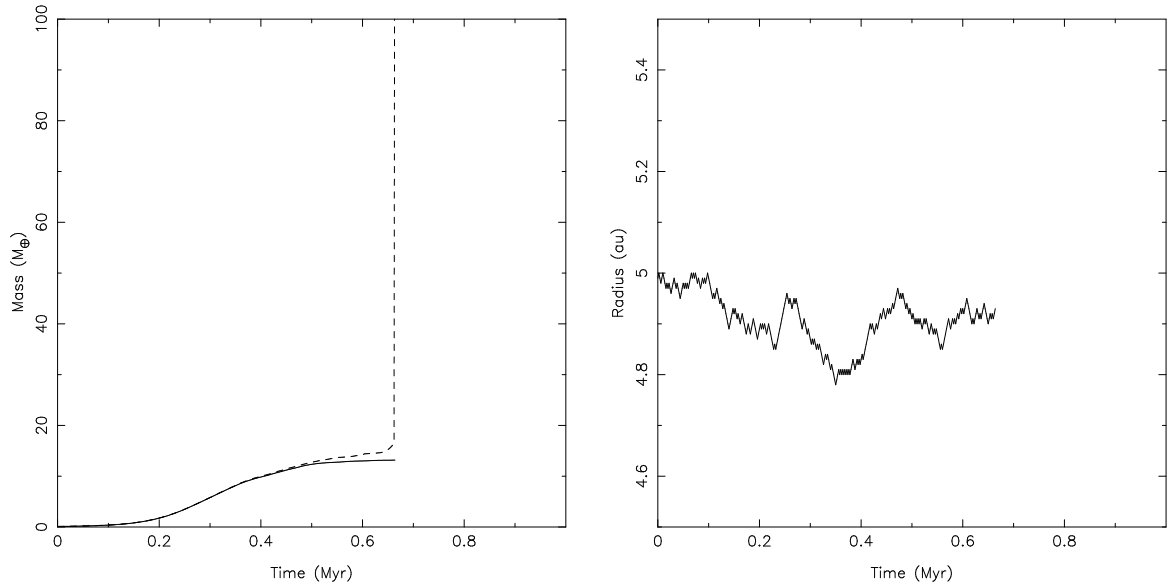}{f2b.eps}
\caption{Left-hand panel: the core mass (solid line) and total planet mass 
(dashed lines) for a core that is allowed to random walk
through the disk. The right-hand panel shows the specific 
migration track for this realization of the model.}
\label{ranmass}
\end{figure}

If the disk at 5~au is turbulent\footnote{There is uncertainty as to 
whether magnetohydrodynamic turbulence can be sustained at radii of a few au, due to 
the low ionization fraction near the disk midplane \citep{gammie96}. The disk may 
only be fully turbulent at late epochs, when the gas 
surface density is low enough that cosmic rays can ionize the interior.}, the latest 
simulations of migration suggest that the core may actually undergo a random walk 
in orbital radius. After $N$ steps of size $\Delta a$, the typical distance moved 
will be $\Delta a \sqrt{N}$. We find that the most interesting behavior occurs 
for planets that wander by a few tenths of an au per Myr, which can be 
accomplished by moving the core either in 
or out by a distance $\Delta a = 0.01 \ {\rm au}$ every $dt = 2000$~yr, with the direction 
of motion determined randomly. This `coarse-grained' approach ignores smaller 
time scale fluctuations, but we have verified that including these (while keeping 
the overall amount of migration constant) does not substantially alter the results. 
Figure \ref{ranmass} shows the core mass and total planet mass as a function of time, 
for a simulation in which the core started at $5$ au, but then  
random walked through the disk. We also plot the migration track of the core 
for this specific realization. The onset of rapid growth of the envelope occurs in a time
of less than $1$ Myr, an order of magnitude less than that obtained for a
core fixed in radius. 

These results can be understood in terms of the relationship between
$M_{\rm core}$ and $M_p$ \citep{papaloizou99}. 
For a given planetesimal accretion rate  
there is a critical core mass $M_{\rm crit}$ above which the core cannot support a stable envelope. 
Below this mass there are generally two solutions, a high-mass solution and a
low-mass solution. For $M_{\rm core} \ll M_{\rm crit}$, the
low-mass solution has a small envelope mass, so that $M_p \approx M_{\rm core}$. 
As the core mass approaches $M_{\rm crit}$, the
envelope mass increases, ultimately contributing $\sim 30$ \% of the total
planet mass. Moreover, the critical mass
increases with increasing planetesimal accretion rate. As a result, 
for a given core mass, the low mass solution decreases with increasing 
accretion rate, while the high mass solution increases with increasing 
accretion rate.

When the core is fixed in radius the planetesimal accretion rate
initially increases with time as the core and planet grow. This growth, however,
depletes the planetesimals in the feeding zone \citep{pollack96}, resulting
in the accretion rate reaching a peak and then declining to an approximately
constant value. For the parameters of the calculation shown in Figure~\ref{Polmass}, 
this plateau is at $7 - 8 \times 10^{-7} M_\oplus \ {\rm yr}^{-1}$. At this
stage the core mass is $\sim 10 M_\oplus$, well below the critical core mass for this
planetesimal accretion rate of $\sim 17 - 18 M_\oplus$ \citep{papaloizou99}. The core
and planet therefore grow slowly, taking $\sim 6 - 7$ Myr, until the core mass
nears the critical core mass. Once at the critical core mass, the low mass
envelope solution is no longer valid. However, since the accretion rate depends
on the total planet mass, a high mass solution is possible. The envelope
mass is therefore determined by the high mass solution and the planet, and 
core, grow rapidly. This halts when the planetesimal accretion rate saturates,
and the total planet mass exceeds the high mass solution for that accretion rate.
This occurs, in this case, at a total planet mass of $\sim 100 M_\oplus$. 

When the core is able to random walk through the disk, the result is quite different. 
The random walk allows the core to not only sample a larger region of the disk, 
but also prevents a single region from being depleted
to the same extent as when the core is stationary. The planetesimal
accretion rate can therefore be maintained at a much higher value, and the core 
grows more rapidly \citep{hourigan84,ward89}. This 
is a generic feature of {\em any} migration model, and it may be 
offset, in part, by the tendency of a migrating core to shepherd 
planetesimals away from the feeding zone \citep{tanaka99}. What is 
more novel about the random walk
case is that the planetesimal
accretion is also highly variable, and this is what ultimately leads to 
the rapid growth phase in this scenario.   

At early times, as in the stationary case, the core mass is 
significantly lower than the local $M_{\rm crit}$, the total 
planet mass is dominated by the core, and the planet grows via
the accretion of planetesimals. In this regime, when 
$M_{\rm core}$ is less than about $10 \ M_\oplus$, the variable 
accretion rate occasioned by migration simply means that the 
planet grows at a variable rate. If, however, the core mass 
approaches the local critical core mass, the envelope mass
can become quite significant. Once $M_{\rm core}$ reaches 
$10 - 12 M_\oplus$, it is possible for the core
to migrate into a region where the accretion rate is low enough
for the core mass to be reasonably close to the local critical
core mass. The planet then acquires an envelope that contributes 
a non-negligible portion of the total planet mass. If, subsequently, 
the core random walks into a region where the accretion rate is 
considerably higher, the low-mass stable envelope solution \citep{papaloizou99} 
would again have the planet mass dominated by the core. This would, 
however, require a decrease in the total planet mass. Such a 
decrease is unlikely to be physical, since it would require 
shedding a bound envelope whose binding energy has presumably 
already been radiated as an additional contributor to the 
planetary luminosity. Accordingly, we assume that the 
low-mass solution remains inaccessible, in which case the 
envelope grows
rapidly, reaching, in this case, a mass of $\sim 200 M_\oplus$. 

Within this model, the core mass at which the rapid growth phase 
occurs depends on the specific random walk of the core. If the core is able to 
stay in a region of high accretion rate for a long time, 
the critical core mass will be higher than if the core, fairly 
early on, random walks into and then out of a region with a
low accretion rate. From additional runs, we find typical 
cores masses are between $12 M_\oplus$ and $15
M_\oplus$. We have also experimented with 
altering the rate of migration. If the planet executes a random 
walk with a larger amplitude -- as some of the simulations 
suggest -- the end result is rapid formation of a gas giant 
with a massive core. For example, a run with $\Delta a = 0.02 \ {\rm au}$ 
and $dt = 10^2 \ {\rm yr}$ led to rapid accretion after 0.5~Myr, 
with a core mass of $25 \ M_\oplus$. We expect qualitatively 
similar results if the migration includes a steady radial drift 
in addition to the fluctuating component, since steady drift 
would similarly allow the core to sample a very wide range of disk radii.

\begin{figure}
\plotone{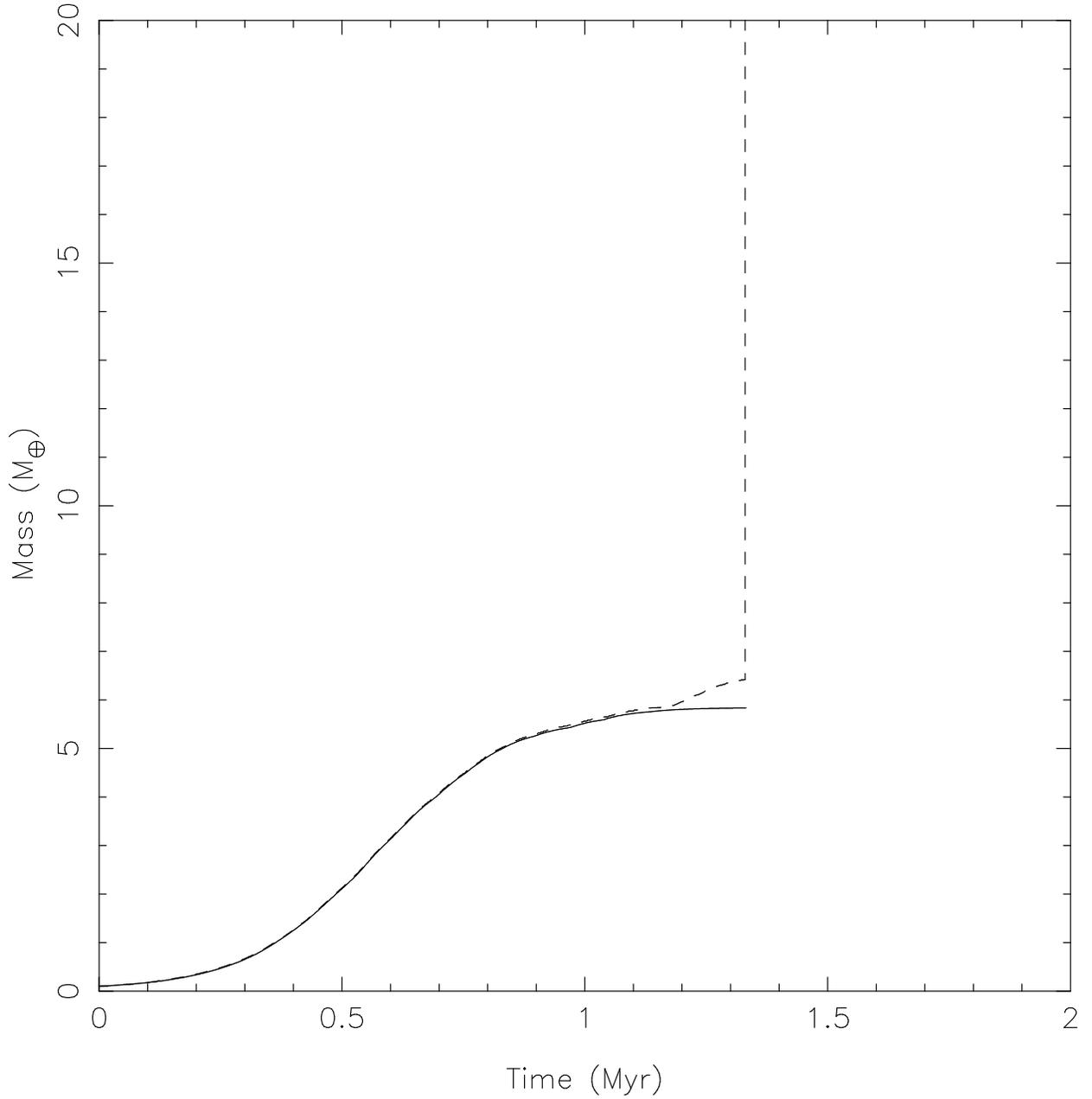}
\caption{Core mass (solid line) and total planet mass
(dashed line) for a core allowed to random walk
through a disk with an initial planetesimal surface 
of $\sigma_p = 5 \ {\rm g \ cm}^{-1}$. Reducing 
$\sigma_p$ leads to a smaller core mass at the 
conclusion of the calculation, while still allowing 
planet formation within the typical lifetime of 
protoplanetary disks.}
\label{jupcore}
\end{figure}

These results suggest that under standard Solar Nebula conditions 
we can form Jupiter rather rapidly, but with a core mass that 
may still be too large \citep{guillot99}. 
Since the derived formation time for Jupiter at 5~au is smaller than the 
lifetime of the Solar Nebula, however, it is possible 
to trade off some of the reduction in formation time for a smaller core 
mass. Figure~\ref{jupcore} shows that 
if we reduce the initial planetesimal 
surface density at 5~au to $5$ g cm$^{-2}$, the lower 
time-averaged accretion rate allows the onset of rapid 
growth to occur at a smaller final core mass. 
In this model the growth time is now longer ($\sim 1.5$ Myr), 
but the core mass at the end of the calculation has been 
reduced to only $\sim 6 M_\oplus$. Depending upon the 
additional mass of heavy elements accreted during the 
subsequent phases of growth, this could be consistent 
with the apparently low core mass of Jupiter. 

\section{Discussion}

If gas giants form within turbulent regions of the protoplanetary disk, 
numerical simulations \citep{nelson03a,laughlin03} show that 
fluctuating disk torques can cause growing cores to wander in orbital 
radius. We have incorporated a simple treatment of this random walk 
migration into models for the formation of gas giants via core 
accretion, and find that for standard Solar Nebula conditions the 
formation time scale of Jupiter can be reduced by almost an 
order of magnitude. Potentially, this could allow massive planets 
to form via core accretion at greater orbital radii, or in disks 
with smaller planetesimal surface densities, than previously suspected. 
For the Solar System, a modestly {\em smaller} 
surface density of planetesimals (5~g~cm$^{-2}$) allows for the timely 
formation of Jupiter with an initial core mass $< 10 M_\oplus$, in better 
agreement with observational constraints. 

For extrasolar planetary systems the main implication is 
that for protoplanetary disks of modest mass and Solar 
metallicity, the formation time scale for giant planets at 5~au is 
significantly less than the typical disk lifetime \citep{haisch01}. At 
smaller radii -- closer to the snow line \citep{sasselov00} -- the 
time scale would be shorter still. This implies that giant planets 
could form in less favorable 
conditions, either in clusters where the disk lifetime was shorter, 
or around lower metallicity stars with smaller reservoirs of 
planetesimals (though if planets are common in 
M4, as suggested by \cite{sigurdsson03}, their 
formation by core accretion is still problematic). 
The observed preponderance of metal-rich stars 
as planetary hosts may then arise from a combination of an 
enhanced {\em probability} of forming multiple planets at 
high metallicity, coupled with frequent destruction of planets 
via Type~II inward migration \citep{armitage02,trilling02}. 

The authors would like to thank John Papaloizou, Richard Nelson and Keith
Horne for useful discussions. WKMR acknowledges support from a PPARC
Standard Grant. This paper is based in part upon work supported 
by NASA under Grant NAG5-13207.

\end{document}